\documentclass[floats,aps,onecolumn,showpacs]{revtex4}
\usepackage{graphicx}
\usepackage{amsmath}
\begin{document}
\newcommand {\nc} {\newcommand}
\nc {\beq} {\begin{eqnarray}} \nc {\eol} {\nonumber \\} \nc {\eeq}
{\end{eqnarray}} \nc {\eeqn} [1] {\label{#1} \end{eqnarray}} \nc
{\eoln} [1] {\label{#1} \\} \nc {\ve} [1] {\mbox{\boldmath $#1$}}
\nc {\rref} [1] {(\ref{#1})} \nc {\Eq} [1] {Eq.~(\ref{#1})} \nc
{\re} [1] {Ref.~\cite{#1}} \nc {\dem} {\mbox{$\frac{1}{2}$}} \nc
{\arrow} [2] {\mbox{$\mathop{\rightarrow}\limits_{#1 \rightarrow
#2}$}}

\author
{E.M. Tursunov}
\address
{Institute of Nuclear Physics, Uzbekistan Academy of Sciences, \\
100214, Ulugbek, Tashkent, Uzbekistan}
\title{A periodic table for the excited $N^*$ and $\Delta^*$ spectrum in a relativistic
chiral quark model }
\date{\today}

\begin{abstract}
 \par A possibility of the construction of a periodic table for
the excited baryon  spectrum is shown in the frame of a relativistic
chiral quark model based on selection rules derived from the
one-pion exchange mechanism. It is shown that all the $N^*$ and
$\Delta^*$ resonances appearing in the $\pi N$ scattering data and
strongly coupling to the $\pi N$ channel are identified with the
orbital configurations $(1S_{1/2})^2(nlj)$. Baryon resonances
corresponding to the orbital configuration with two valence quarks
in excited states couple strongly to the $\pi \pi N$-channel, but
not to the $\pi N$ channel.
 \par At low energy scale up to 2000 MeV, the obtained numerical estimations
 for the SU(2) baryon states (up to and including F-wave
 $N^*$ and $\Delta^*$ resonances) within  the schematic periodic
 table are mostly consistent with the experimental data.
 It is argued that due-to the overestimation of the ground state $N(939)$ and Roper
resonance $N(1440)$ almost by the same amount and that the Roper
resonance is a radial excitation of the $N(939)$, the "lowering
mechanism" for the both baryon states must be the same. The same
mechanism is expected in the $\Delta$ sector. At higher energies,
where the experimental data are poor, we can extend our model
schematically and  predict seven new $N^*$ and four $\Delta^*$
resonances with larger spin values.
\end{abstract}


\maketitle

\section{Introduction}

\par It is believed during the long period that the baryon spectrum
can be described with a good accuracy in the Constituent Quark
Models (CQM) based either on the Goldstone-boson exchange (GBE)
\cite{glo98}, or one-gluon exchange (OGE) \cite{cap86} or (and)
instanton induced exchange (IIE) \cite{met01} mechanisms between
non-relativistic constituent quarks. However, there are still some
serious problems which cannot be avoided until now. The most
important issue is the problem of "missing resonances": the CQMs
predicted too many 
states even at low energies which are not observed at the
experimental facilities \cite{PDG,rev,anis}. The situation is so
serious, that "a modern view questions the usefulness of quarks to
describe the nucleon excitation spectrum" \cite{anis}.
\par Fortunately, there are also strong optimistic views on the problem.
 The recent review \cite{rev} shows all the difficulties in baryon
spectroscopy and concluded that the all photo-, pion- and
hadron-induced reactions will be important to understand the excited
baryon spectroscopy. From the theoretical side, essential
developments are being done in the Lattce QCD \cite{lat1,lat2},
Dyson-Schwinger equations \cite{roberts}, effective
field-theoretical methods \cite{obu11,mul10,lue10} and in the
theoretical coupled channel approaches to meson-baryon scattering
within the Juelich \cite{krewald} and Sato-Lee \cite{lee} models.

\par In \cite{tur05,tur09,tur10} we have developed a relativistic
chiral quark model \cite{gutsche89} for the lower excitation
spectrum of the nucleon and delta. For the first time the mass
values of the Roper resonance and lowest negative parity nucleon and
delta resonances have been estimated in a relativistic chiral quark
model. The splitting of the Roper resonance from the N(939) was
reproduced with a reasonable accuracy.
\par The aim of this paper is to show that in a way to extend the
relativistic chiral quark model to the higher excitation spectrum,
it is possible to construct a periodic table for the excited $N^*$
and $\Delta^*$ states based on the one-pion exchange mechanism
between valence quarks. In the model the $SU(2)$-baryons (Nucleon
and $\Delta$ states) are assumed to correspond to the orbital
configuration $(1S_{1/2})^2(nlj)$ with two S-valence quarks and a
single valence quark in the excited state. We will prove that such
baryon states are identified with all the $N^*$ and $\Delta^*$
resonances appearing in the $\pi N$ scattering process. Then we can
compare our theoretical construction with the experimental data. Our
analysis is common for all relativistic chiral quark models
describing the baryons as bound states of three valence quarks with
a Dirac two-component structure and surrounded by the cloud of
$\pi$-mesons, as required by the chiral symmetry
\cite{tur05,tur09,tur10,gutsche89,thomas80,thomas08,saito84,saito08,waka10}.

\section{Model}
\par The effective Lagrangian of our model
${\cal L}(x)$ contains the quark core part ${\cal L}_Q(x)$, the
quark-pion  ${\cal L}_I^{(q\pi)}(x)$  interaction part, and the
kinetic parts for the pion ${\cal L}_{\pi}(x)$ :
\begin{eqnarray}
\nonumber
{\cal L}(x) = {\cal L}_Q(x) + {\cal L}_I^{(q\pi)}(x) + {\cal L}_{\pi}(x) \\
 \nonumber
 = \bar\psi(x)[i\not\!\partial -S(r)-\gamma^0V(r)]\psi(x) \\
 -  1/f_{\pi}
 \bar\psi[S(r) i \gamma^5 \tau^i \phi_i]\psi+ \frac{1}{2}[(\!\partial_{\mu}\phi_i)^2
 -m_{\pi} \phi_i^2 ].
\end{eqnarray}
Here, $\psi(x)$ and $\phi_i, i=1,2,3$ are the quark and pion fields,
respectively. The matrices $\tau^i (i=1,2,3)$ are for the isospin.
The pion decay constant $f_\pi=$93 MeV. The scalar part of the
static confinement potential is given by
\begin{equation}
S(r)=cr+m
\end{equation}
where c and m are constants.

\par At short distances, transverse fluctuations of the string are dominating,
 with some indication that they transform like the time component
of the Lorentz vector. They are given by a Coulomb type vector
potential as
\begin{equation}
\label{Coulomb}
 V(r)=-\alpha/r
\end{equation}
where $\alpha$ is approximated by a constant. The quark fields are
obtained from solving the Dirac equation with the corresponding
scalar plus vector potentials
\begin{equation} \label{Dirac}
[i\gamma^{\mu}\partial_{\mu} -S(r)-\gamma^0V(r)]\psi(x)=0
\end{equation}
The respective positive and negative energy eigenstates as solutions
to the Dirac equation with a spherically symmetric mean field, are
given in a general form as
\begin{eqnarray} \label{Gaussian_Ansatz}
 u_{\alpha}(x) \, = \,
\left(
\begin{array}{c}
g^+_{N\kappa }(r) \\
-i f^+_{N\kappa }(r) \,\vec{\sigma}\hat{\vec x} \\
\end{array}
\right) \, {\cal Y}_{\kappa}^{m_j}(\hat{\vec x}) \,\chi_{m_t} \,
\chi_{m_c} \, e^{-iE_{\alpha}t}
\end{eqnarray}

\begin{eqnarray}
 v_{\beta}(x) \, = \,
\left(
\begin{array}{c}
g^-_{N\kappa}(r) \\
-i f^-_{N\kappa}(r) \,\vec{\sigma}\hat{\vec x} \\
\end{array}
\right) \, {\cal Y}_{\kappa}^{m_j}(\hat{\vec x}) \,\chi_{m_t} \,
\chi_{m_c} \, e^{+iE_{\beta}t}
\end{eqnarray}
The quark and anti-quark eigenstates $u$ and $v$ are labeled by the
radial, angular, azimuthal, isospin and color quantum numbers $N,\,
\kappa,\, m_j,\, m_t$ and $m_c$, which are collectively denoted by
$\alpha$ and $\beta$, respectively. The spin-angular part of the
quark field operators
\begin{equation}
{\cal Y}_{\kappa}^{m_j}(\hat{\vec x})\,=\,[Y_l(\hat{\vec x})\otimes
\chi_{1/2}]_{jm_j} \, \, j=|\kappa|-1/2.
\end{equation}
The quark fields $\psi$ are expanded over the basis of positive and
negative energy eigenstates as
\begin{equation}
\psi(x)=\sum \limits_{\alpha} u_{\alpha}(x)b_{\alpha} +\sum
\limits_{\beta} v_{\beta}(x)d^{\dag}_{\beta} .
\end{equation}
The expansion coefficients $b_{\alpha}$ and $d^{\dag}_{\beta}$ are
operators, which annihilate a quark and create an anti-quark in the
orbits $\alpha$ and $\beta$, respectively.
\par The free pion field operator is expanded over plane wave solutions as
\begin{equation}
\phi_j(x)=(2\pi)^{-3/2}\,
\int\frac{d^3k}{(2\omega_k)^{1/2}}[a_{j{\bf
k}}exp(-ikx)+a^{\dag}_{j{\bf k}}exp(ikx)]
\end{equation}
with the usual destruction and creation operators $a_{j{\bf k}}$ and
$a^{\dag}_{j{\bf k}}$ respectively. The pion energy is defined as \\
$\omega_k \,=\, \sqrt{k^2+m_{\pi}^2}. $ The expansion of the free
zero mass gluon field operators is of the same form.

\par Using the effective Lagrangian and the time-ordered perturbation theory
 one can develop a calculation scheme for the excitation spectrum of the nucleon and delta.
 At the zero-th order, the quark core result ($E_Q$) is obtained by solving Eq.(\ref{Dirac}) for the
single quark system numerically. The corresponding quark core energy
is evaluated as the sum of single quark energies:
$$ E_Q=2E(1S_{1/2}) + E(nlj)$$ with an appropriate correction on the
center of mass motion \cite{tur09}. Since we work in the independent
particle model, the bare three-quark state of the $SU(2)$-flavor
baryons is assumed to correspond to the orbital structure
$(1S_{1/2})^2(nlj)$ in the non-relativistic spectroscopic notation.
This assumption is very natural, since an orbital or radial
excitation of two or three valence quarks in the baryon state must
be suppressed strongly. However, we remember that a single valence
quark state is described with the two-component Dirac wave function.
The orbital momenta corresponding to these two components differ by
one unity. Further we will see that the two-component Dirac
structure of the quark wave function plays a crucial role in the
derivation of the selection rules for the baryon state total
momentum.
 \par The second order perturbative corrections to the energy spectrum of the
 SU(2) baryons due to the pion field ($\Delta E^{(\pi)}$) are calculated on
 the basis of the Gell-Mann and Low theorem :

 \begin{eqnarray}\label{Energy_shift}
\hspace*{-.8cm} \nonumber \Delta E=<\Phi_0| \,
\sum\limits_{i=1}^{\infty} \frac{(-i)^n}{n!} \,   \int \,
i\delta(t_1) \, d^4x_1 \ldots d^4x_n \, T[{\cal H}_I(x_1) \ldots
{\cal H}_I(x_n)] \, |\Phi_0>_{c}
\end{eqnarray}
with $n=2$, where the relevant quark-pion interaction Hamiltonian
density is
\begin{eqnarray}
{\cal H}_I^{(q\pi)}(x)= \frac{i}{f_{\pi}}\bar\psi(x)\gamma^5
\vec\tau\vec\phi(x)S(r)\psi(x),
\end{eqnarray}

 The stationary bare three-quark state $|\Phi_0>$ is constructed
from the vacuum state using the usual creation operators:
\begin{equation}
|\Phi_0>_{\alpha\beta\gamma}=b_{\alpha}^+b_{\beta}^+b_{\gamma}^+|0>,
\end{equation}
where $\alpha, \beta$ and $ \gamma$ represent the quantum numbers of
the single quark states, which are coupled to the respective baryon
configuration. The energy shift of Eq.(\ref{Energy_shift}) is
evaluated up to second order in the quark-pion interaction, and
generates self-energy and exchange diagrams contributions.

\par The self-energy terms contain contribution both from
intermediate quark $(E>0)$ and anti-quark $(E<0)$ states. These
diagrams correspond to the case when a single pion is emitted and
absorbed by the same valence quark which is excited to the
intermediate quark and anti-quark states. The convergence of the
self-energy diagrams was shown explicitly for the lowest $1S$, $2S$,
$1P_{1/2}$ and $1P_{3/2}$ valence quark states \cite{tur10}.

\par The second-order one-pion exchange diagrams at one loop yield an
additional correction to the mass spectrum of the SU(2) baryon
state. The estimation of the lowest excitation spectrum of the
Nucleon and $\Delta$ in Ref. \cite{tur09} shows that is is possible
to reproduce the main properties of the excited baryon spectrum
based on the one-pion loop corrections.

\section{Selection rules for the quantum numbers of the excited
 $N^*$ and $\Delta^*$ states}
\par Now we begin to analysis the excited
$N^*$ and $\Delta^*$ spectrum based on the relativistic description
of the one-pion exchange mechanism (see \cite{tur09}). We do not
write down here explicitly the expression for the pion-exchange
operator, but only remember that it couples the upper and lower
components of the two interacting valence quarks, respectively. In
this way we can derive the selection rules for the quantum numbers
of the baryon states with the fixed orbital configuration.
\par Let us to fix the orbital configuration as $(1S_{1/2})^2(nlj)$,
with the intermediate spin coupling $\vec{S}_0=\vec{S}_1+\vec{S}_2 =
\vec{1/2}+\vec{1/2}$ of the two $1S$-valence quarks, where the last
valence quark $(nlj)$ can be in the ground or an excited state. The
upper and lower Dirac components of the last excited valence quark
have orbital momenta $l$ and $l' =l\pm 1$, respectively. It is clear
that these configurations must describe well the lowest baryon
excited states. However, the question is, how successful  are they
in higher excitation spectrum? Our choice of the above orbital
configuration is close to the limitation in the diquark-quark models
\cite{diquark}, where some of the degrees of freedom are "frozen".
The corresponding baryon states are different from members of the
SU(6)$\otimes$O(3) multiplets in the Constituent Quark Models.

\par The first two selection rules come from the coupling of the three
valence quarks into the SU(2) baryon state with total momentum $J$
and isospin $T$:
\begin{eqnarray}
\label{quarkcoup}
 \nonumber
\vec{S}_0+\vec{j}=\vec{J} \\
 \vec{S}_0+\vec{1/2} =\vec{T},
\end{eqnarray}
where the symmetry property of the two S-quark coupling was used.
The third rule comes from the pion exchange mechanism between the
excited valence quark and the $1S$ quark. This mechanism couples the
upper (lower) component of the $1S$ valence quark with the lower
(upper) component of the excited $(nlj)$ valence quark. Since the
upper component of the S-quark has zero orbital momentum, then for
the orbital momentum of the exchanged pion we derive the equation
\begin{equation}
\label{piorbit}
 L_{\pi}=l'=l \pm 1
\end{equation}

\par   The final selection rule is based on the assumption that the coupling of
the last valence quark with quantum numbers $(nlj)$ to the $1S$
quark plus pion  is the main component of the strong coupling of the
excited baryon state to the $N(939)+\pi$:
\begin{equation}
\label{piN}
 \vec{L}_{\pi}+\vec{1/2} =\vec{J}
\end{equation}
With this assumption, Eq.(\ref{piorbit}) can be used for the
identification of the baryon resonance in the $\pi N$-scattering
process. Namely, when $l'=0$ we have S-wave nucleon and delta
resonances, when $l'=1$ we have P-wave resonances, etc.
\par An important consequence of the obtained selection rules is
that all the $N^*$ and $\Delta^*$ resonances appearing in the $\pi
N$ scattering process and strongly coupling to the $\pi N$ channel
are identified with the orbital configurations $(1S_{1/2})^2(nlj)$
with two valence quarks in the ground state and a single valence
quark in an excited state. A baryon resonance corresponding to the
orbital configuration with two valence quarks in excited states
$(1S_{1/2})(nlj)_1(nlj)_2$ couples strongly to the $\pi \pi
N$-channel, but not to the $\pi N$ channel.

\par Using the obtained selection rules it is very natural to
analysis the excited nucleon and delta spectrum. For the fixed
orbital configurations $(1S_{1/2})^2(nlj)$ with the intermediate
spin coupling of the two $S-$ wave quarks $S_0=0$ (the so-called
instanton channel), Eq.(\ref{quarkcoup}) allows only a single $N^*$
state with $J=j$ and no any $\Delta^*$ resonances.
\par Except the case, when the last valence quark is in the $P_{1/2}$ orbit,
the intermediate coupling $S_0=1$, due-to the selection rule
Eq.(\ref{piN}) yields two resonances in the both nucleon and delta
sectors with the total momentum $J=L_{\pi} \pm 1/2$. In this way one
of the $N^*$ resonances defined by the selection rules in
Eq.(\ref{quarkcoup}) with $J=j+1$ or $J=j-1$ is ruled out. When the
last valence quark is in the $P_{1/2}$ orbit, i.e. has the lower
S-component, the selection rules yield $L_{\pi}=0$ and $J=1/2$, and
consequently, only single S-wave resonances in the both nucleon and
$\Delta$ sectors are allowed.
\par Thus, for the fixed $(1S_{1/2})^2(nlj)$  orbital configuration
with $(nlj) \ne (nP_{1/2})$ there must be three $N^*$ and two
$\Delta^*$ resonances. The lightest $N^*$ state corresponds to the
intermediate spin coupling $S_0=0$ due-to strong attraction in this
"instanton channel". The other two $N^*$, as well as the two
$\Delta^*$ resonances correspond to the spin coupling $S_0=1$ and
must be close each to other.
\par In the case when the last quark is
in the $P_{1/2}$ orbit, there are two $N^*$ states (not close each
to other) and a single $\Delta^*$ resonance appearing in the S-wave
of the $\pi N$ scattering.

\section{Numerical results}

\par In Table 1 we compare our numerical estimations of the excited
$N^*$ and $\Delta^*$ spectrum within the developed schematic
periodic  table with the last experimental data from \cite{rev} and
\cite{anis}. The calculations were done up-to and including F-wave
baryon resonances in the frame of the developed chiral quark model.
Details of the calculations were given in \cite{tur05,tur09,tur10}.
In the Table we give the center of mass (CM) corrected quark core
results (zero order estimation) (second column)  together with the
second order pion field contributions corresponding to the self
energy (3-th column) and exchange diagrams (second order
corrections) (4-th column). The final estimations are given in the
5-column with the fixed value of the weak decay coupling constant
$f_{\pi}=93$ MeV. The results in the 6-column were obtained with the
renormalized coupling constant $f'_{\pi}(ex)=65.22$ MeV for the
exchange diagrams, while keeping the standart value $f_{\pi}=93$ MeV
for the self-energy terms.

We first fix the orbital configuration $(1S_{1/2})^2(nS_{1/2})$. In
the data there are four $N^*$ with $J^{\pi}=1/2^+$ ( $P_{11}$
resonances) and two $N^*$ with $J^{\pi}=3/2^+$ ( $P_{13}$
resonances). With the above rules, we can find easily that
$N^*(1440)$,  $N^*(1710)$, and $N^*(1720)$ resonances belong to the
orbital configuration $(1S_{1/2})^2 (2S_{1/2})$ with the radially
excited $2S$ valence quark state, while the other three $N^*(1880)$,
$N^*(1900)$ and $N^*(2100)$ resonances correspond to the orbital
configuration
 $(1S_{1/2})^2(3S_{1/2})$. In the $\Delta$ sector there are two
resonances with $J^{\pi}=3/2 ^+$  at  1600 MeV and 1920 MeV, and two
states with $J^{\pi}=1/2 ^+$ at 1750 MeV and 1910 MeV which belong
to the orbital configuration with the radially excited valence quark
in consistence with our results.
\par The orbital configuration $(1S_{1/2})^2(1D_{3/2})$ is not
presented in the data, since it would give two $N^*$ resonances with
$J^{\pi}=3/2^+$ and a single  $N^*$ resonance with $J^{\pi}=1/2^+$.

\par For the orbital configurations $(1S_{1/2})^2(nP_{1/2})$ there
are four nucleon and three delta resonances with $J^{\pi}=1/2 ^-$
and they are not close each to others. Each of the nucleon bands
$n=1$ and $n=2$  contains two resonances, while $\Delta^*$
resonances correspond to the three bands including $n=3$.
\par The orbital configuration $(1S_{1/2})^2(nP_{3/2})$ with $n=1$
 yields three $N^*$ resonances $3/2^-$(1520), $5/2^-$(1675) and $3/2^-$(1700),
 the first of which is less than other two states in accordance with our prediction.
The band with $n=2$ yields next group of the D-wave Nucleon
resonances  $3/2^-$(1860), $3/2^-$(2080) and $5/2^-$(2200).
 \par In the Delta sector there are four D-wave resonances, however
 only two of them $\Delta(5/2^-)$(1930) and  $\Delta(3/2^-)$(1940)
 are close each to other. Since other D-wave resonances
$\Delta(3/2^-)$(1700) and  $\Delta(5/2^-)$(2350) are far, then we
can predict possible new  $\Delta^*(5/2^-)$ (around 1700 MeV) and
$\Delta^*(3/2^-)$ (around 2350 MeV) resonances.
\par The F-wave $N^*$ resonances $N^*(5/2^+)(1680)$,  \\
$N^*(5/2^+)(1870)$  and  $N^*(7/2^+)(1990)$ belong to the orbital
configuration $(1S_{1/2})^2(nD_{5/2})$ with $n=1$ together with
delta states $\Delta^*(5/2^+)$(1905) and $\Delta^*(7/2^+)$(1950),
while the $\Delta^*(5/2^+)$(2000) and $\Delta^*(7/2^+) $(2390)
belong to the $n=2$ band.
\par We can continue our analysis at higher energies and predict
in summary seven new $N^*$  resonances  with $J^{\pi}=7/2^-$ ~ (2000
MeV), $9/2^+$ ~(2100 - 2300 MeV),~
 $11/2^+$ (2100 - 2300 MeV),~ $11/2^-$ ~(2500-2700 MeV),~ $13/2^-$ (2500-2700 MeV),
 ~$13/2^+$ ~(2600 -2800 MeV), ~$15/2^+$ ~(2600 -2800 MeV) and four  $\Delta^*$
 resonances with $J^{\pi}=5/2^-$~ (around 1700 MeV),~ $3/2^-$ ~(2350 MeV),
 ~$11/2^-$ ~(2750 MeV),~ $13/2^+$ ~(2950 MeV). These resonances are expected to be observed
 in current experimental facilities.
\par It is clear now that the remaining "missing $N^*$ and $\Delta^*$ resonances"
predicted by the Constituent Quark Models must appear in the $\pi
\pi N$ strong coupling sector, if they exist. As we have argued
above, they will be assigned with the orbital configuration
$(1S_{1/2})(nlj)_1(nlj)_2$ with two excited valence quarks and a
single ground state valence quark.
\par  Now we can analysis the numerical values in Table 1 in comparison
with the experimental data-analysis from Ref. \cite{rev}. As can be
seen from the Table, the mass spectrum of the Nucleon and $\Delta$
is described reasonably well in the relativistic chiral quark model.
The important observation is that one needs an additional exchange
mechanism for the lowering both the ground state N(939) and Roper
resonance N(1440) almost by the same amount. This fact indicate that
the "lowering mechanism" for the both N(939) and Roper resonance
$N^*(1440)(1/2^+)$ should be the same, since these states have
identical quantum numbers except the radial quantum number. The same
situation is in the $\Delta$ sector. Additionally, most of the
radially excited $N^*$, except $N^*(1710)(1/2^+)$ and
$N^*(2100)(1/2^+)$ and all the radially excited $\Delta^*$
resonances are overestimated in our model. The exception is possible
due-to the experimental errors.
\par Contrary, the orbitally excited $N^*$ resonances are
mostly underestimated, except the states $N^*(1650)(1/2^-)$,
$N^*(1860)(3/2^-)$ and $N^*(1905)(1/2^-)$. The situation in the
$\Delta^* $ sector is different. The negative parity $\Delta^*$
states are consistent with the experimental data: exception is here
for the $\Delta^*(1700)(3/2^-)$, which is underestimated by 70 - 170
MeV. The positive parity $\Delta^*(1905)(5/2^+)$ and
$\Delta^*(1950)(7/2^+)$ resonances are also underestimated by some
amount.
\par The use of the renormalized coupling constant
$f'_{\pi}(ex)=65.22$ MeV for the exchange diagrams yields a good
estimation for energy values of the ground states $N(939)$ and
$\Delta(1232)$ and their energy difference, but strongly moves down
the lowest orbitally excited baryon resonances $N(1520)$, $N(1535)$,
$\Delta(1620)$ and $\Delta(1700)$.
\par The analysis shows that one needs an additional exchange
mechanism between valence quarks to reproduce the whole SU(2) baryon
spectrum. The new exchange forces  must depend on the spin and
flavor of valence quarks as well as on the quantum numbers of the
baryon state. Of course, some part of the interaction comes from
gluon fields and two-pion exchange mechanism.

\section{Conclusions}
 \par In summary, we have derived selection rules for the excited baryon state, assuming
 that it's orbital configuration is of the form $(1S)^2(nlj)$ with two valence
 quarks in the ground state and a single excited quark. These selection rules were derived
 on the basis of the one-pion exchange mechanism between valence
 quarks in the frame of the relativistic chiral quark model.
 An important consequence of the obtained selection rules is that all
the $N^*$ and $\Delta^*$ resonances appearing in the $\pi N$
scattering process and strongly coupling to the $\pi N$ channel are
identified with the orbital configurations $(1S_{1/2})^2(nlj)$.
 Baryon resonances corresponding to the orbital configuration with
two valence quarks in excited states couple strongly to the $\pi \pi
N$-channel, but not to the $\pi N$ channel.
\par Based on obtained selection rules, we have constructed a schematic periodic table
for the excited $N^*$ and $\Delta^*$ spectrum. The obtained
numerical estimations for the energy positions of baryon resonances
(up to and including F-wave) are consistent with the experimental
data.
\par The important observation is that one needs an additional exchange
mechanism for the lowering both the ground state $N(939)$ and Roper
resonance $N(1440)$ almost by the same amount. This fact indicate
that the "lowering mechanism" for the both N(939) and Roper
resonance $N^*(1440)(1/2^+)$ should be the same, since these states
have identical quantum numbers except the radial quantum number. The
same situation is in the $\Delta$ sector.

\par At higher energies, where the experimental data are poor, we can
extend our model schematically and  predict seven new $N^*$ and four
$\Delta^*$ states with larger spin values. Of course, the number of
"missing resonances" in our model is strongly suppressed due-to
restriction of the configuration space to the orbits
$(1S_{1/2})^2(nlj)$. However, as we have shown above, at lower
energies this construction works reasonably well.

\section*{Acknowledgments}
 Author thanks S. Krewald for valuable and very useful discussions.

\begin{table}
\caption { Estimations for the energy values of the  $N^*$ and
$\Delta^*$ states in MeV  up to and including F-wave resonances with
the CM correction in the LHO method}
 \begin{tabular}{|c|c|c|c|c|c|c|}     \hline
   & $E_Q (CM corrected) $   &  $ \Delta E_{\pi}(s.e.)$ &  $\Delta E_{\pi}(ex.)$ &
    $E_Q+\Delta E_{\pi}$ & $E_Q+\Delta E'_{\pi}$ & exp.\cite{rev}  \\  \hline

$N(939)(1/2^+)$ $(1S)^3$  & 966 & 380 & -180 & 1166 & 980 & 938 $\div$ 939    \\
\hline

$N(1440)(1/2^+)$ $(1S)^2(2S)$ & 1200 & 603 &-113 & 1690& 1573 & 1430 $\div$ 1470  \\
\hline
$N(1710)(1/2^+)$ $(1S)^2(2S)$ & 1200 & 603 &-66 & 1737 &1669 & 1700 $\div$ 1750  \\
\hline
$N(1720)(3/2^+)$ $(1S)^2(2S)$ & 1200 & 603 & 1 & 1804& 1805  & 1700 $\div$ 1760  \\
\hline
$N(1880)(1/2^+)$ $(1S)^2(3S)$ & 1361 & 788 &-110 & 2039& 1925 & 1840 $\div$ 1940  \\
\hline
$N(2100)(1/2^+)$ $(1S)^2(3S)$ & 1361 & 788 &-39 & 2110& 2070  & 2000 $\div$ 2200  \\
\hline
$N(1900)(3/2^+)$ $(1S)^2(3S)$ & 1361 & 788 &-3  & 2146& 2143 & 1900 $\div$ 2000  \\
\hline

 $N(1535)(1/2^-)$ $(1S)^2 1P_{1/2}$ & 1129&501&-119&1511& 1388 &1528$\div$1548
 \\  \hline
$N(1650)(1/2^-)$ $(1S)^2 1P_{1/2}$ & 1129&501&46 &1676 & 1724
&1640$\div$1680
 \\  \hline

$N(1905)(1/2^-)$ $(1S)^2 2P_{1/2}$&1301&713&-111&1903&
1788&1850$\div$1950
  \\  \hline
 $N(2090)(1/2^-)$ $(1S)^2 2P_{1/2}$ & 1301&713&24&2038 & 2063&  2100$\div$2260
 \\  \hline

 $N(1520)(3/2^-)$ $(1S)^2 1P_{3/2}$ & 1107 &515 & -126 & 1496 & 1366  &1518 $\div$ 1526  \\  \hline
$N(1700)(3/2^-)$ $(1S)^2 1P_{3/2}$ & 1107 &515 & -79 & 1543 & 1461 &
1675 $\div$ 1775  \\  \hline
$N(1675)(5/2^-)$ $(1S)^2 1P_{3/2}$ & 1107 &515 & 11 & 1633 & 1645  & 1670 $\div$ 1680 \\
\hline

$N(1860)(3/2^-)$ $(1S)^2 2P_{3/2}$ & 1293 &713 & -111 & 1895& 1781
&1810 $\div$ 1890  \\  \hline

$N(2080)(3/2^-)$ $(1S)^2 2P_{3/2}$ & 1293 &713 & -31 & 1975& 1943
&2045 $\div$ 2155  \\  \hline

$N(2070)(5/2^-)$
$(1S)^2 2P_{3/2}$ & 1293 &713 & 4 & 2010 & 2014 &2075 $\div$ 2245  \\
\hline

$N(1680)(5/2^+)$
$(1S)^2 1D_{5/2}$ & 1212 &638 & -114 & 1736 & 1618 &1680 $\div$ 1690  \\
\hline

$N(1870)(5/2^+)$
$(1S)^2 1D_{5/2}$ & 1212 &638 & -37 & 1813& 1775 &1840 $\div$ 1960  \\
\hline

$N(1990)(7/2^+)$
$(1S)^2 1D_{5/2}$ & 1212 &638 & 12 & 1862 & 1874 &1860 $\div$ 2100  \\
\hline

$ \Delta(1232)(3/2^+)$  $(1S)^3$  & 966 & 380 & -36 & 1310 & 1273
&1230 $\div$ 1234  \\  \hline

$ \Delta(1600)(3/2^+)$ $(1S)^2(2S)$ & 1200 & 603 & -23 &1780& 1756 &
1535 $\div$ 1695  \\  \hline
 $ \Delta(1750)(1/2^+)$ $(1S)^2(2S)$ & 1200
& 603 & 1 &1804 & 1805 & 1710 $\div$ 1780  \\  \hline

$\Delta(1910)(1/2^+)$ $(1S)^2(3S)$ & 1361 & 788 & -3 &2146 & 2143 &
1845 $\div$ 2025  \\  \hline

 $\Delta(1920)(3/2^+)$ $(1S)^2(3S)$ & 1361 & 788 & -18 &2131 & 2112 & 1880
$\div$ 2020  \\  \hline

$\Delta(1620)(1/2^-)$  $(1S)^2 1P_{1/2}$ & 1129&501& -24 &1606 &
1581& 1603 $\div$ 1649  \\
\hline
 $\Delta(1900)(1/2^-)$  $(1S)^2 2P_{1/2}$ & 1301&713&  -24 &
 1990& 1965 & 1860 $\div$ 1960  \\  \hline

$\Delta(1700)(3/2^-)$ $(1S)^2 1P_{3/2}$ & 1107 &515 &-18 & 1604 &
1585& 1670 $\div$ 1770   \\  \hline
  $\Delta(5/2^-)$ $(1S)^2 1P_{3/2}$ & 1107 &515 &-35 & 1587 & 1551 & ...   \\
\hline

$\Delta(1940)(3/2^-)$ $(1S)^2 2P_{3/2}$ & 1293 &713 &-9 & 1997 &
1988& 1935 $\div$ 2055   \\  \hline
 $\Delta(1930)(5/2^-)$ $(1S)^2 2P_{3/2}$ & 1293 &713 &-22 & 1984 & 1961 & 1900 $\div$ 1960   \\
\hline

$\Delta(1905)(5/2^+)$
$(1S)^2 1D_{5/2}$ & 1212 &638 & -12 & 1838 & 1826 & 1860 $\div$ 1940  \\
\hline

$\Delta(1950)(7/2^+)$
$(1S)^2 1D_{5/2}$ & 1212 &638 & -27 & 1823 & 1795 &1915 $\div$ 1960  \\
\hline

\end{tabular}
\end{table}

\end{document}